\pdfoutput=1

\documentclass[reprint,aps,prl,twocolumn,,floatfix,superscriptaddress]{revtex4-1}
\usepackage{graphicx}
\usepackage{dcolumn}
\usepackage{bm}
\usepackage{amsmath}
\usepackage{lineno}
\usepackage[english]{babel}
\usepackage[utf8]{inputenc}
\usepackage{color}
\usepackage{siunitx}
\usepackage{amsmath}

\begin{document}

\title{Wafer-scale low-disorder 2DEG in $^{28}$Si/SiGe without an epitaxial Si cap}

\author{Davide Degli Esposti}
\affiliation{QuTech and Kavli Institute of Nanoscience, Delft University of Technology, Lorentzweg 1, 2628 CJ Delft, Netherlands}
\author{Brian Paquelet Wuetz}
\affiliation{QuTech and Kavli Institute of Nanoscience, Delft University of Technology, Lorentzweg 1, 2628 CJ Delft, Netherlands}
\author{Viviana Fezzi}
\affiliation{QuTech and Kavli Institute of Nanoscience, Delft University of Technology, Lorentzweg 1, 2628 CJ Delft, Netherlands}
\author{Mario Lodari}
\affiliation{QuTech and Kavli Institute of Nanoscience, Delft University of Technology, Lorentzweg 1, 2628 CJ Delft, Netherlands}
\author{Amir Sammak}
\affiliation{QuTech and Netherlands Organisation for Applied Scientific Research (TNO), Stieltjesweg 1, 2628 CK Delft, The Netherlands}
\author{Giordano Scappucci}
\email{g.scappucci@tudelft.nl}
\affiliation{QuTech and Kavli Institute of Nanoscience, Delft University of Technology, Lorentzweg 1, 2628 CJ Delft, Netherlands}

\date{\today}

\begin{abstract}
We grow $^{28}$Si/SiGe heterostructures by reduced-pressure chemical vapor deposition and terminate the stack without an epitaxial Si cap but with an amorphous Si-rich layer obtained by exposing the SiGe barrier to dichlorosilane at 500~$^{\circ}$C. As a result, $^{28}$Si/SiGe heterostructure field-effect transistors feature a sharp semiconductor/dielectric interface and support a two-dimensional electron gas with enhanced and more uniform transport properties across a 100~mm wafer. At $T=1.7$~K we measure a high mean mobility of $(1.8 \pm 0.5) \times 10^5$ cm$^2$/Vs and a low mean percolation density of $(9 \pm 1) \times 10^{10}$~cm$^{-2}$. From the analysis of Shubnikov--de~Haas oscillations at $T=190$~mK we obtain a long mean single particle relaxation time of $(8.1 \pm 0.5)$~ps, corresponding to a mean quantum mobility and quantum level broadening of $(7.5 \pm 0.6) \times 10^4$~cm$^{2}$/Vs and $(40\pm3)$~$\si{\micro\electronvolt}$, respectively, and a small mean Dingle ratio of $(2.3\pm0.2)$, indicating reduced scattering from long range impurities and a low-disorder environment for hosting high-performance spin-qubits. 
\end{abstract}

\maketitle

Strained $^{28}$Si/SiGe heterostructures are a compelling platform for scalable qubit tiles based on gate-defined quantum dots.\cite{Vandersypen2017InterfacingCoherent,Lieaar3960} In these $^{28}$Si buried quantum wells, electron spins experience a quiet electrical and magnetic environment. The electronically noisy semiconductor/dielectric interface is far away, separated from the quantum well by a SiGe epitaxial barrier, and the nuclear spins have been removed by isotopic enrichment. Continuous advances in the material science of $^{28}$Si/SiGe and improved device fabrication have enabled quantum logic with spin qubits crossing the surface code threshold,\cite{xue2022quantum,noiri2022fast,mills2021two} coherent coupling of two electron spins at a distance via virtual microwave photons,\cite{harvey-collard_circuit_2021} and CMOS-based cryogenic control of quantum circuits\cite{xue2021cmos}. In the mainstream approach to quantum dot fabrication, the last step in the heterostructure growth cycle comprises the heteroepitaxial deposition of a thin epitaxial Si cap on the SiGe barrier.\cite{lawrie2020quantum} This is to avoid the formation of low-quality Ge-based oxides upon exposure of SiGe to air. After the Si cap deposition, a high-$\kappa$ dielectric is deposited \textit{ex}-\textit{situ} and at low-temperature ($\approx 300$~$^{\circ}$C) to insulate the gate from the buried and undoped quantum well. This low-temperature process preserves the strain in the quantum well but induces large concentrations of impurities at the critical semiconductor/dielectric interface. These impurities can influence the electrostatic confining potential landscape induced by the gates, leading to the formation of unintentional quantum dots,\cite{thorbeck2015formation} and are a source of charge noise limiting qubit performance.\cite{connors2019low,struck_lowfrequency_2020} While efforts have focused on achieving uniform and high-purity $^{28}$Si quantum wells with sharp interfaces,\cite{wuetz2021atomic,chen_detuning_2021,hollmann_large_2020} now more attention is needed to optimize the step which terminates the heterostructure deposition cycle and has a critical role in defining the semiconductor/dielectric interface. 

\begin{figure}[t]
	\includegraphics[width=83mm]{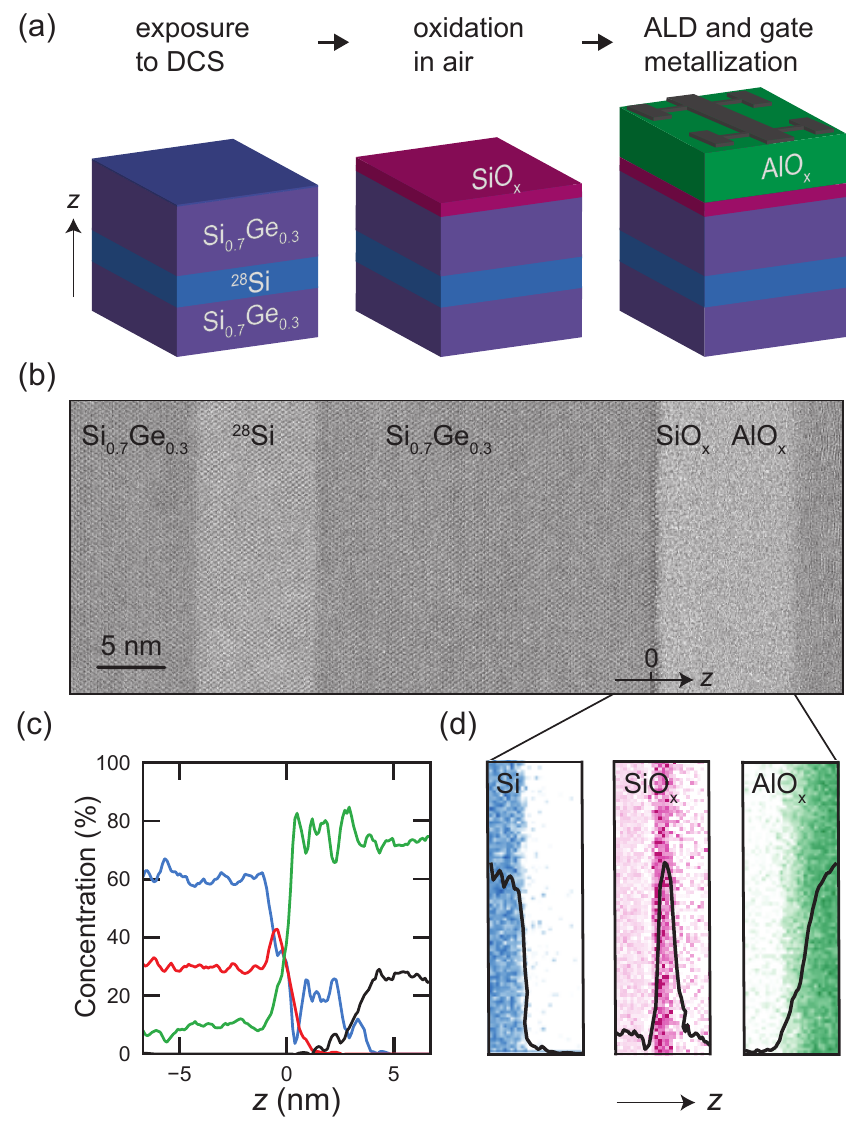}
	\caption{(a) Schematics of the $^{28}$Si/SiGe heterostructure and formation of the dielectric interface in a Hall-bar heterostructure field effect transistor.~$z$ indicates the heterostructure growth direction. The heterostructure is terminated by exposure to dichlorosilane (DCS) gas at a temperature below the threshold for growing an epitaxial Si cap and the dielectric stack comprises a SiO$_x$ layer formed by exposure of the heterostructure to air at room temperature and an AlO$_x$ layer formed by atomic layer deposition (ALD).
	(b) BF-STEM image of the active layers of the $^{28}$Si/SiGe heterostructure field effect transistor showing, from left to right, the Si$_{0.7}$Ge$_{0.3}$ strain-relaxed buffer layer, the tensile-strained $^{28}$Si quantum well, the Si$_{0.7}$Ge$_{0.3}$ barrier, followed by the SiO$_x$/AlO$_x$ dielectric stack. 
	(c) Electron energy loss spectroscopy (EELS) semi-quantitative concentration depth profiles across the semiconductor/dielectric interface for Si (blue), Ge (red), O (green), and Al (black).
	(d) 15~nm$\times$45~nm wide 2D maps by EELS using low-energy edges to recognize differences between the different bonding states: Si (blue), SiO$_x$ (magenta), and AlO$_x$ (green). We do not detect any Cl or H signal above the background noise in our EELS data.}

\label{fig:one}
\end{figure}

In this letter, we explore $^{28}$Si/SiGe heterostructures terminated by exposure to dichlorosilane (DCS) gas at a temperature well below the threshold for epitaxial growth of Si. By avoiding the growth of an epitaxial Si cap altogether, we obtain $^{28}$Si/SiGe heterostructure field effect transistors (H-FETs) with a sharp semiconductor/dielectric interface. We show that the $^{28}$Si quantum well supports a two-dimensional electron gas with less disorder and improved quantum transport properties compared to heterostructures with an epitaxial Si cap.

Figure~\ref{fig:one}(a) illustrates the workflow to fabricate $^{28}$Si/SiGe H-FETs. We grow $^{28}$Si/SiGe heterostructures on 100~mm Si(001) wafers using an Epsilon 2000 (ASMI) reduced-pressure chemical vapor deposition reactor. 
We use isotopically-enriched $^{28}$SiH$_4$ for growing the $^{28}$Si quantum well (residual $^{29}$Si concentration of 0.08\%\cite{SabbaghPhysRevApplied2019,xue2021cmos,xue2022quantum}) and DCS (H$_2$SiCl$_2$) and GeH$_4$ for all other layers. 
The heterostructure comprises a $3~\si{\micro\meter}$ step-graded Si$_{1-x}$Ge$_x$ layer (final $x \simeq 0.3$), a $2.5~\si{\micro\meter}$ Si$_{0.7}$Ge$_{0.3}$ strain-relaxed buffer, a 8~nm tensile-strained $^{28}$Si quantum well and a 30~nm Si$_{0.7}$Ge$_{0.3}$ barrier\footnote{A typical secondary ions mass spectrometry of our heterostructures is reported in Fig.~S13 of Ref.~\cite{wuetz2021atomic}. The oxygen concentration in the $^{28}$Si quantum well is $\simeq 4 \times 10^{17}$~cm$^{-3}$} and in the SiGe barrier is $\simeq 4 \times 10^{17}$~cm$^{-3}$. To achieve sharp interfaces and minimize Si/Ge interdiffusion at the quantum well-barrier interface\cite{wuetz2021atomic}, the temperature is decreased from 750~$^{\circ}$C for growing the quantum well to 625~$^{\circ}$C for the barrier. We now introduce a major difference compared to our previous experiments. In Refs.~\cite{samkharadze2018strong,xue2021cmos,xue2022quantum,wuetz2021atomic}) we deposited a thin epitaxial Si cap at 675~$^{\circ}$C using DCS. Here we reduce the substrate temperature to 500~$^{\circ}$C, below the desorption temperature of chlorine from the surface (600--650~$^{\circ}$C),\cite{gao_chlorine_1993,bolotov_electronic_2021} under the same conditions of DCS flow and pressure. According to literature\cite{hartmann_reduced_2008,hartmann_epitaxial_2009, bauer_novel_2010, vincent_low_2010, vincent_si_2011, hartmann_benchmarking_2012,aubin_epitaxial_2016}, we expect that exposure to DCS at 500~$^{\circ}$C essentially suppresses crystalline growth but creates an amorphous Si-rich layer on Si$_{0.7}$Ge$_{0.3}$.
After terminating the deposition cycle with this step, the heterostructure is removed from the growth reactor and a native oxide is formed upon exposure to air at room temperature. We identify the native oxide as SiO$_x$ based on the chemical analysis in Fig.~\ref{fig:one}(c),(d). Then, we fabricate Hall-bar shaped H-FETs using the process described in Ref.~\cite{wuetz2021atomic}. In short, the process comprises the implantation of ohmic contacts and rapid thermal annealing at 700~$^{\circ}$C, the atomic layer deposition at 300~$^{\circ}$C of a 10~nm~Al$_2$O$_3$ dielectric layer on the SiO$_x$, and the final deposition of a Hall-bar shaped metallic gate, electrically insulated from the heterostructure by the SiO$_x$/Al$_2$O$_3$ dielectric stack.

Figure~\ref{fig:one}(b) shows a bright-field scanning transmission electron microscopy (BF-STEM) image of the heterostructure and of the dielectric stack under the gate stack at the end of the H-FET fabrication process. The Si quantum well is uniform, without extended defects, and is characterized by sharp top and bottom interfaces to the Si$_{0.7}$Ge$_{0.3}$ layers, in agreement with our previous reports.\cite{xue2021cmos,xue2022quantum,wuetz2021atomic} The semiconductor/dielectric interface is similarly sharp, highlighted by the perfect atomically sharp semiconductor surface as imaged by BF-STEM. Two distinct amorphous layers, which we identify as the SiO$_x$ and AlO$_x$ layers, appear on the dielectric side of the interface. We gain insights over the nature of the semiconductor/dielectric interface and of the dielectric stack by performing electron energy loss spectroscopy (EELS)(Supplementary). In Fig.~\ref{fig:one}(c) we show the semi-quantitative concentration profiles using the Si-K (1839-2084 eV), Al-K (1560-1700 eV), O-L (532-660 eV), and Ge-L (1220-1400 eV) high energy edge. The Si (blue) and Ge (red) concentration profiles decrease together whilst the oxygen (green) signal is increasing. We deduce that oxidation of the Si$_{0.7}$Ge$_{0.3}$ barrier with on top an amorphous Si-rich layer results in a sharp SiGe/SiO$_x$ semiconductor/dielectric interface. This is confirmed by the minor Ge pile-up on the semiconductor side of the interface,\cite{liu1995hydridation,long_ge_2012} which appears as a dark line in BF-STEM [Fig.~\ref{fig:one}(b)] and suggests that the top of the single crystalline Si${_0.7}$Ge$_{0.3}$ barrier has been oxidized and that Ge oxides at the interface are absent\cite{liou1991effects,legoues1989oxidation}. Furthermore, the Al signal (black line) rises after the Si signal from SiO$_x$ has trailed, indicating that the dielectric stack retains the two distinct SiO$_x$ and AlO$_x$ layers. 

In Fig.~\ref{fig:one}(d) we show the chemical mapping by EELS of Si (blue), SiO$_x$ (magenta), and AlO$_x$ (green) along and across the semiconductor/dielectric interface, together with the intensity profiles. To recognize differences between the different bonding states, we use the low-energy Si-L edge (96.3-100.8 eV) for the semiconductor phase and a shifted Si-L edge (101.4-107.1 eV) for the oxide phase, and Al-L (73.8-79.5 eV) for the oxided Al phase. The SiGe/SiO$_x$ interface is sharp throughout the image, whereas the SiO$_x$/AlO$_x$ interface shows some interdiffusion. By fitting the intensity profiles with exponential functions\cite{schubert1994delta} we characterize the size of the interfaces with the leading (towards the surface) and trailing (from the surface) exponential slopes $\lambda_L$ and $\lambda_T$. We find $\lambda_L^{\text{Si}} = (1.0 \pm 0.1)$~nm and $\lambda_T^{\text{SiO}_\text{x}} = (0.8\pm0.1)$~nm. Conversely, we find $\lambda_L^{\text{SiO}_\text{x}} = (1.9\pm0.1)$ nm and $\lambda_T^{\text{AlO}_\text{x}}= (3.1\pm0.2)$ nm. Overall, the transition from epitaxial SiGe to amorphous SiO$_x$ interface is sharper than the transition between SiO$_x$ and AlO$_x$, pointing to a degree of intermixing at the latter interface.

We characterized the H-FETs by magnetotransport measurements at a temperature of 1.7~K and 190~mK\footnote{$T = 190$~mK is the electron temperature obtained by fitting Coulomb blockade peaks (Supplementary) measured on quantum dot devices\cite{xue2021cmos} fabricated on a similar heterostructure. The electron temperature is higher than the temperature of 70~mK measured by a thermometer located on the mixing chamber of the dilution refrigerator} in refrigerators equipped with cryo-multiplexers.\cite{wuetz2019multiplexed} With this approach, we measure multiple devices from a wafer in the same cool-down. The devices are operated in accumulation mode, in which electrons populate the undoped $^{28}$Si quantum well by applying a positive DC gate voltage ($V_G$). We measure the longitudinal and transverse components of the resistivity tensor, $\rho_{xx}$ and $\rho_{xy}$, by using standard four-probe lock-in techniques at fixed AC source-drain bias of 100~$\si{\micro\volt}$. We calculate the longitudinal $\sigma_{xx}$ and transverse $\sigma_{xy}$ conductivity via tensor inversion. We measure electron density ($n$) and mobility ($\mu$) with the classical Hall effect at low perpendicular magnetic field $B$.

\begin{figure}[t]
	\includegraphics[width=83mm]{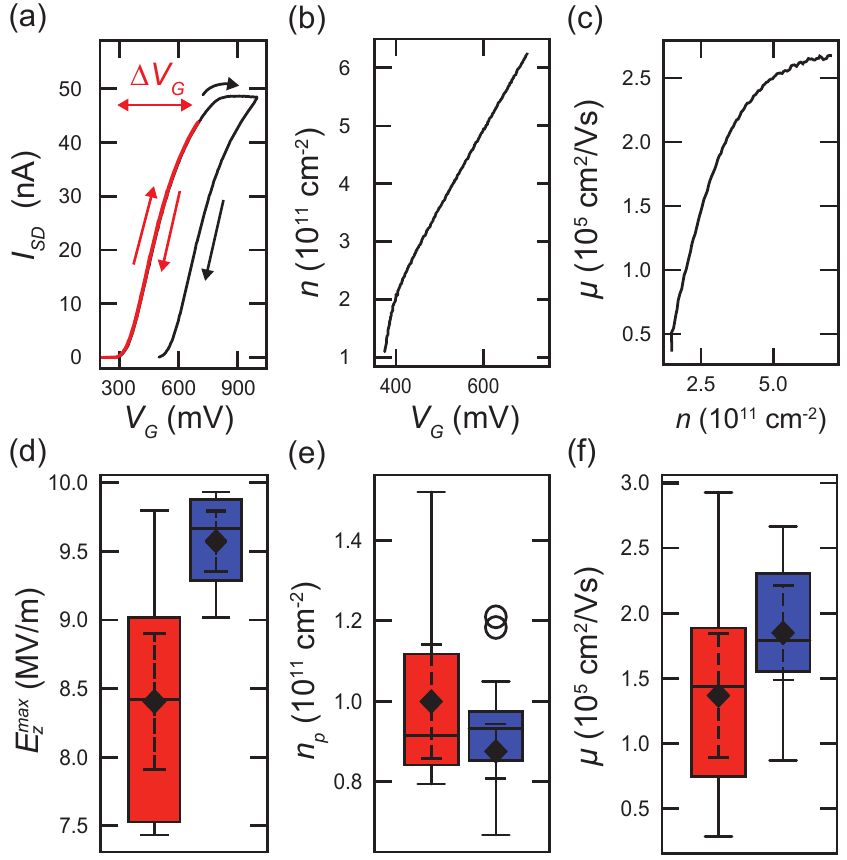}%
	\caption{
	(a) Source-drain current $I_{SD}$ measured at $T = 1.7$~K as a function of gate voltage $V_G$ for a typical Hall bar heterostructure field effect transistor (H-FET). The operational gate voltage range $\Delta V_{G}$ indicates the range over which an $I_{SD}$-$V_{G}$ curve (red line) can be measured repeatedly without hysteresis and drift. 
	(b) Density $n$ as a function of gate voltage $V_G$ and (c) electron mobility $\mu$ as a function of $n$ measured within the operational gate voltage range.
	(d), (e), (f) Distributions of maximum electric field applicable before hysteresis $E_z^{max}$, percolation density $n_p$, and $\mu$ measured at $n = 6 \times 10^{11}$ cm$^{-2}$ for heterostructures terminated by a Si-rich amorphous layer obtained exposure to DCS at 500~$^{\circ}$C (blue, 14 H-FETs measured) and for heterostructures with an epitaxial Si cap grown by exposure to DCS at 675~$^{\circ}$C (red, 16 H-FETs measured). Quartile box plots, mode (horizontal line), means (diamonds), outliers (circles), and $99$\% confidence intervals of the mean (dashed whiskers) are shown.}
\label{fig:two}
\end{figure}

Figure~\ref{fig:two}(a) shows for a typical device the turn-on and pinch-off source-drain current $I_{SD}$ as a function of increasing and decreasing $V_G$, respectively. Above a threshold voltage ($V_G = 350$~mV), the current starts flowing in the channel and increases monotonically. If the gate voltage is operated within the operational gate voltage range $\Delta V_G$ (red curve), $I_{SD}$ is stable and the threshold and pinch-off voltages overlap. At higher $V_G$, $I_{SD}$ saturates due to charge build-up at the semiconductor/dielectric interface, triggering hysteresis and, consequently, a shift in pinch-off voltage. As shown in Fig.~\ref{fig:two}(b), if $V_G$ is swept within the operational gate voltage range, $n$ increases linearly with $V_G$ up to $6\times 10^{11}$~cm$^{-2}$. From the slope $\frac{dn}{dV_G}$ we derive an effective capacitance per unit area $C \simeq 205$~nF/cm$^2$ using the relationship $C = e\frac{dn}{dV_G}$.\cite{wuetz2019multiplexed}. This capacitance characterizes the parallel-plate capacitor where the 2DEG in the $^{28}$Si quantum well and the metallic top gate are insulated by a SiGe/SiO$_x$/AlO$_x$ dielectric stack. Figure~\ref{fig:two}(c) shows the density-dependent mobility measured in the same density range as in Fig.~\ref{fig:two}(b).
In the low density regime ($n \leq 3\times 10^{11}$~cm$^{-2}$), the mobility rises steeply due to the increasing screening of Coulomb scattering from remote charged impurities located at semiconductor/dielectric interface.\cite{monroe_comparison_1993} At higher density ($n \geq 5\times 10^{11}$~cm$^{-2}$), the mobility approaches saturation at a value above $2.5 \times 10^5$~cm$^2$/Vs. This weaker density-dependence is typical of a high-quality 2DEG, where the maximum mobility is limited by short-range scattering from impurities within or near the quantum well.\cite{mi2015magnetotransport,laroche_scattering_2015,wuetz2019multiplexed}

In Fig.~\ref{fig:two}(d)--(f) we plot the distributions of the maximum electric field ($E_z^{max}$), the percolation density ($n_p$), and the mobility at high density for heterostructures terminated with an amorphous Si-rich layer (blue) and, as a benchmark, for heterostructures with an epitaxial Si cap (red). These three metrics are obtained from the analysis of measurements in Fig.~\ref{fig:two}(a)--(c), repeated on multiple H-FETs on dies that are randomly selected from different locations across the 100~mm~wafer. $E_z^{max}$, calculated as ${C \Delta V_G}/{\epsilon_0 \epsilon_r}$, where $\epsilon_r=11.68$ is the dielectric constant of Si, indicates the maximum electric field that we can apply to the quantum well in the H-FETs before hysteresis. Large $E_z^{max}$ are desirable for device stability, increased tunability, and large valley splitting.\cite{Friesen2007ValleyWells,wuetz2020effect,hollmann_large_2020,wuetz2021atomic} $n_p$ characterizes disorder in low density regime, relevant for quantum dot operation, and is obtained by fitting the density-dependent $\sigma_{xx}$ to percolation theory.\cite{Tracy2009ObservationMOSFET} Finally, the mobility at high density is a probe for disorder arising from within or nearby the quantum well.\cite{monroe_comparison_1993,laroche_scattering_2015,mi2015magnetotransport} Overall, H-FETs perform better when the SiGe barrier is terminated with an amorphous Si-rich layer. We measure a 9\% increase in mean $E_z^{max}$, 7\% decrease in mean percolation density, and a 40\% increase in mean mobility. Most importantly, we observe a reduction in the spread of $E_z^{max}$, $n_p$, and $\mu$ of $\simeq 300 \%$, $\simeq 200 \%$, and $\simeq 30 \%$ respectively, pointing to an increased uniformity on a 100~mm wafer scale. 

\begin{figure}[hhh!]
	\includegraphics[width=86mm]{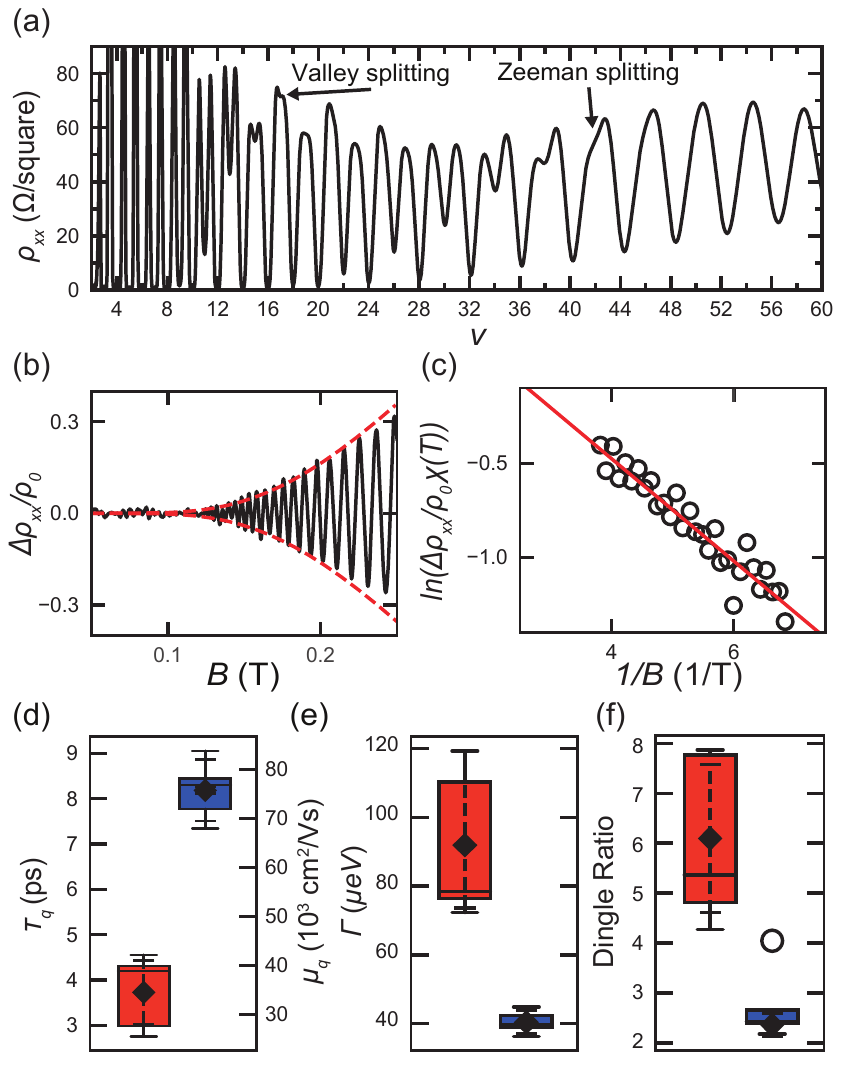}%
	\caption{(a) Longitudinal resistivity $\rho_{xx}$ measured at $T = 190$~mK as a function of Landau level filling factor $\nu$. 
	These measurements are performed at fixed $n = 4.75 \times 10^{11}$ cm$^{-2}$ while sweeping the perpendicular magnetic field $B$. Spin and valley degenerate Landau levels correspond to $\nu$ = 4$k$ ($k$ = 1,2,3...), Zeeman split levels to $\nu$ = (4$k$-2), whereas valley split levels correspond to odd integer filling factors $\nu$. 
	Arrows indicate the filling factors at which Zeeman spin splitting and valley splitting are resolved.
	(b) Normalized resistivity oscillation amplitude (black curve) as a function of $B$ after polynomial background subtraction. The arrow indicates the magnetic field at which Shubnikov--de~Haas oscillations are resolved. The red dashed line is the theoretical fit of the oscillations envelope from which we extract $\tau_q$. 
	(c) Dingle plot (open circles) from the first twenty most resolved resistivity oscillation maxima and minima and theoretical curve (solid red line) computed using $\tau_q$ from the analysis in (b).
	(d), (e), (f) Distributions of $\tau_q$, $\mu_q$, $\Gamma$, and Dingle ratio measured at $n = (5-6)\times 10^{11}$ cm$^{-2}$ for heterostructures terminated by a Si-rich amorphous layer obtained exposure to DCS at 500~$^{\circ}$C (blue, 5 H-FETs measured) and for heterostructures with an epitaxial Si cap grown by exposure to DCS at 675~$^{\circ}$C (red, 7 H-FETs measured). Quartile box plots, mode (horizontal line), means (diamonds), outliers (circles), and $99$\% confidence intervals of the mean (dashed whiskers) are shown.}

\label{fig:three}
\end{figure}

We further characterize disorder in the $^{28}$Si/SiGe heterostructure at 190~mK by measuring the single-particle relaxation time $\tau_q$\cite{das_sarma_single-particle_1985} in the quantum Hall regime. From $\tau_q$ we derive the quantum mobility $\mu_q = e\tau_q/m^*$, where $e$ is the elementary charge and $m^*$ is the effective mass, and the quantum level broadening of the momentum eigenstates $\Gamma = \hbar/2\tau_q$, here $\hbar$ is the reduced Planck constant. $\mu_q$, associated with $\tau_q$, is influenced by all scattering events and is different from the mobility $\mu = e\tau_t/m^*$, where the scattering time $\tau_t$ is unaffected by forward scattering. Therefore $\tau_q$ and $\mu_q$ qualify the disorder in the heterostructure more comprehensively than $\tau_t$ and $\mu$. 

Figure~\ref{fig:three}(a) shows for the H-FET with the highest mobility a measurement of $\rho_{xx}$ plotted for clarity against the Landau level filling factor $\nu = h n/ e B$, where $h$ is the Plank constant. This measurement was performed at fixed density $n = 4.75 \times 10^{11}$ cm$^{-2}$ by keeping $V_G$ constant and sweeping $B$. Onset of Shubnikov--de Haas oscillation, Zeeman splitting, and valley splitting occurs at $0.125$, $0.43$, and $1.15$~T, respectively, corresponding to $\nu = 152$, $42$ and $17$. The observation of Shubnikov--de~Haas oscillations, Zeeman and valley splitting at these high filling factors indicates a very low level of disorder.\cite{qian2017quantum} Figure~\ref{fig:three}(b) shows the normalized oscillation amplitude $\Delta \rho_{xx}/\rho_0 = (\rho_{xx}- \rho_0)/\rho_0$ in the low magnetic field regime after polynomial background subtraction. $\rho_0 \simeq 63$~$\Omega$/square is the longitudinal resistivity at zero magnetic field from which we extract a mobility of $2.7 \times 10^5$ cm$^2$/Vs. We estimate $\tau_q = (7.4\pm0.1)~\text{ps}$ from a fit of the Shubnikov-de Haas oscillation envelope to the function $\Delta\rho_{xx} = 4 \rho_0 \chi(T) \exp(-\pi / \omega_c \tau_q)$, where $\chi(T) = (2\pi^2 k_B T/\hbar\omega_c)/ \sinh{(2\pi^2 k_B T/\hbar\omega_c)}$. Here $T=190$~mK, $k_B$ is the Boltzmann constant, and $\omega_c$ is the cyclotron frequency calculated using a fixed $m^* = 0.19\,m_e$.\cite{coleridge1991small,qian2017quantum} From $\tau_q$ we derive $\mu_q = (6.8\pm0.1) \times 10^4$~cm$^{2}$/Vs, $\Gamma = (44\pm1)$~$\si{\micro\electronvolt}$, and find a Dingle ratio $\tau_t/\tau_q\simeq 3.8$. The Dingle plot of Fig.~\ref{fig:three}(c) highlights the high number of oscillation maxima and minima used in the fitting procedure.

In Fig.~\ref{fig:three}(d)--(f) we plot the distributions for $\tau_q$ (and $\mu_q$), $\Gamma$, and the Dingle ratio $\tau_t/\tau_q$, measured in the high density regime ($n = (5-6) \times 10^{11}$) cm$^{-2}$). As in Fig.~\ref{fig:two}(d)--(f), we consider heterostructures terminated with an amorphous Si-rich layer (blue, 5 H-FETs measured) and heterostructures with an epitaxial Si cap (red, 7 H-FETs measured). Heterostructures with an amorphous Si-rich layer have a mean $\tau_q$ of $(8.1\pm0.5)$ ps, and consequently a mean $\mu_q$ of $(7.5 \pm 0.6) \times 10^4$~cm$^{2}$/Vs and $\Gamma$ of $(40\pm3)$~$\si{\micro\electronvolt}$, representing a $\simeq 2\times$ improvement compared to heterostructures with an epitaxial Si cap. Consistent with the trend in Fig.~\ref{fig:two}(d)--(f), we find a significant reduction in spread for $\tau_q$ (30\%), and consequently for $\mu_q$, $\Gamma$. Furthermore, in heterostructures with an amorphous Si-rich layer we find a mean Dingle ratio of $(2.3\pm0.2)$. This mean value is $\simeq 300\%$ smaller and has an 80\% reduction in spread compared to heterostructures with an epitaxial Si cap. This low value of the Dingle ratio indicates that short-range scattering from impurities within or near the quantum well is the dominant scattering mechanism\cite{monroe_comparison_1993}, in agreement with the analysis of the mobility-density curve. Scattering from remote impurities is reduced thanks to a better semiconductor/dielectric interface. Our mean value for $\tau_q$ in $^{28}$Si/SiGe is also on par with the best value reported in Ref.~\cite{mi2015magnetotransport} from H-FETs in Si/SiGe heterostructures featuring an epitaxial Si cap. However, in our samples, the semiconductor/dielectric interface is much closer to the channel (30~nm compared to 50~nm in Ref.~\cite{mi2015magnetotransport}). Therefore, this comparison confirms that scattering from remote impurities is limited in our devices as a consequence of a high-quality and uniform semiconductor/dielectric interface associated with the termination process at 500~$^{\circ}$C.

In summary, we challenged the mainstream approach to deposit an epitaxial Si cap on $^{28}$Si/SiGe heterostructures and, instead, we terminated the SiGe barrier with an amorphous Si-rich layer, obtained by exposure to DCS at 500~$^{\circ}$C. Compared to previous heterostructures that feature an epitaxial Si cap and that have already produced high performance spin qubits,\cite{xue2021cmos,xue2022quantum}, we demonstrate an improvement in performance of H-FETs in terms of mean value and spread of mobility, percolation density, maximum electric field before hysteresis, and single particle relaxation time (and hence quantum mobility). We speculate that performance improves because the amorphous Si-rich layer gets completely oxidized compared to the epitaxial Si cap (Supplementary), thereby creating a more uniform SiO$_x$ layer with less scattering centers. By having a better semiconductor/dielectric interface and wafer-scale uniformity, we expect that this material stack may lead to Si spin qubits with improved yield and performance. In this direction, charge noise measured in quantum dots on these heterostructures will be very informative as these measurements probe the dynamics of charge fluctuations that transport experiments are not very sensitive to. These results motivate new studies, for example by varying the temperature and/or time of exposure to DCS to understand in detail the nature of the amorphous Si-rich layer on the SiGe barrier, the role of Cl and H upon oxidation in air, and to use this knowledge as a tool for further optimizing the semiconductor/dielectric interface.

\vspace{\baselineskip}
This research was supported by the European Union’s Horizon 2020 research and innovation programme under the grant agreements No. 951852 (QLSI project), and in part by the Army Research Office (Grant No. W911NF-17-1-0274). 
The views and conclusions contained in this document are those of the authors and should not be interpreted as representing the official policies, either expressed or implied, of the Army Research Office (ARO), or the U.S. Government. 
The U.S. Government is authorized to reproduce and distribute reprints for Government purposes notwithstanding any copyright notation herein. 
This work is part of the research program OTP with project number 16278, which is (partly) financed by the Netherlands Organisation for Scientific Research (NWO).

\vspace{\baselineskip}
Data sets supporting the findings of this study are available at \url{https://doi.org/10.4121/19181597}.

\section{Supplemental Material}
Supplemental material is provided with an extended Fig.~\ref{fig:one} and measurements of the electron temperature. 

\bibliography{bibliography.bib}

\begin{thebibliography}{42}%
\makeatletter
\providecommand \@ifxundefined [1]{%
 \@ifx{#1\undefined}
}%
\providecommand \@ifnum [1]{%
 \ifnum #1\expandafter \@firstoftwo
 \else \expandafter \@secondoftwo
 \fi
}%
\providecommand \@ifx [1]{%
 \ifx #1\expandafter \@firstoftwo
 \else \expandafter \@secondoftwo
 \fi
}%
\providecommand \natexlab [1]{#1}%
\providecommand \enquote  [1]{``#1''}%
\providecommand \bibnamefont  [1]{#1}%
\providecommand \bibfnamefont [1]{#1}%
\providecommand \citenamefont [1]{#1}%
\providecommand \href@noop [0]{\@secondoftwo}%
\providecommand \href [0]{\begingroup \@sanitize@url \@href}%
\providecommand \@href[1]{\@@startlink{#1}\@@href}%
\providecommand \@@href[1]{\endgroup#1\@@endlink}%
\providecommand \@sanitize@url [0]{\catcode `\\12\catcode `\$12\catcode
  `\&12\catcode `\#12\catcode `\^12\catcode `\_12\catcode `\%12\relax}%
\providecommand \@@startlink[1]{}%
\providecommand \@@endlink[0]{}%
\providecommand \url  [0]{\begingroup\@sanitize@url \@url }%
\providecommand \@url [1]{\endgroup\@href {#1}{\urlprefix }}%
\providecommand \urlprefix  [0]{URL }%
\providecommand \Eprint [0]{\href }%
\providecommand \doibase [0]{http://dx.doi.org/}%
\providecommand \selectlanguage [0]{\@gobble}%
\providecommand \bibinfo  [0]{\@secondoftwo}%
\providecommand \bibfield  [0]{\@secondoftwo}%
\providecommand \translation [1]{[#1]}%
\providecommand \BibitemOpen [0]{}%
\providecommand \bibitemStop [0]{}%
\providecommand \bibitemNoStop [0]{.\EOS\space}%
\providecommand \EOS [0]{\spacefactor3000\relax}%
\providecommand \BibitemShut  [1]{\csname bibitem#1\endcsname}%
\let\auto@bib@innerbib\@empty
%</preamble>
\bibitem [{\citenamefont {Vandersypen}\ \emph {et~al.}(2017)\citenamefont
  {Vandersypen}, \citenamefont {Bluhm}, \citenamefont {Clarke}, \citenamefont
  {Dzurak}, \citenamefont {Ishihara}, \citenamefont {Morello}, \citenamefont
  {Reilly}, \citenamefont {Schreiber},\ and\ \citenamefont
  {Veldhorst}}]{Vandersypen2017InterfacingCoherent}%
  \BibitemOpen
  \bibfield  {author} {\bibinfo {author} {\bibfnamefont {L.~M.~K.}\
  \bibnamefont {Vandersypen}}, \bibinfo {author} {\bibfnamefont
  {H.}~\bibnamefont {Bluhm}}, \bibinfo {author} {\bibfnamefont {J.~S.}\
  \bibnamefont {Clarke}}, \bibinfo {author} {\bibfnamefont {A.~S.}\
  \bibnamefont {Dzurak}}, \bibinfo {author} {\bibfnamefont {R.}~\bibnamefont
  {Ishihara}}, \bibinfo {author} {\bibfnamefont {A.}~\bibnamefont {Morello}},
  \bibinfo {author} {\bibfnamefont {D.~J.}\ \bibnamefont {Reilly}}, \bibinfo
  {author} {\bibfnamefont {L.~R.}\ \bibnamefont {Schreiber}}, \ and\ \bibinfo
  {author} {\bibfnamefont {M.}~\bibnamefont {Veldhorst}},\ }\href {\doibase
  10.1038/s41534-017-0038-y} {\bibfield  {journal} {\bibinfo  {journal} {npj
  Quantum Information}\ }\textbf {\bibinfo {volume} {3}},\ \bibinfo {pages}
  {34} (\bibinfo {year} {2017})}\BibitemShut {NoStop}%
\bibitem [{\citenamefont {Li}\ \emph {et~al.}(2018)\citenamefont {Li},
  \citenamefont {Petit}, \citenamefont {Franke}, \citenamefont {Dehollain},
  \citenamefont {Helsen}, \citenamefont {Steudtner}, \citenamefont {Thomas},
  \citenamefont {Yoscovits}, \citenamefont {Singh}, \citenamefont {Wehner},
  \citenamefont {Vandersypen}, \citenamefont {Clarke},\ and\ \citenamefont
  {Veldhorst}}]{Lieaar3960}%
  \BibitemOpen
  \bibfield  {author} {\bibinfo {author} {\bibfnamefont {R.}~\bibnamefont
  {Li}}, \bibinfo {author} {\bibfnamefont {L.}~\bibnamefont {Petit}}, \bibinfo
  {author} {\bibfnamefont {D.~P.}\ \bibnamefont {Franke}}, \bibinfo {author}
  {\bibfnamefont {J.~P.}\ \bibnamefont {Dehollain}}, \bibinfo {author}
  {\bibfnamefont {J.}~\bibnamefont {Helsen}}, \bibinfo {author} {\bibfnamefont
  {M.}~\bibnamefont {Steudtner}}, \bibinfo {author} {\bibfnamefont {N.~K.}\
  \bibnamefont {Thomas}}, \bibinfo {author} {\bibfnamefont {Z.~R.}\
  \bibnamefont {Yoscovits}}, \bibinfo {author} {\bibfnamefont {K.~J.}\
  \bibnamefont {Singh}}, \bibinfo {author} {\bibfnamefont {S.}~\bibnamefont
  {Wehner}}, \bibinfo {author} {\bibfnamefont {L.~M.~K.}\ \bibnamefont
  {Vandersypen}}, \bibinfo {author} {\bibfnamefont {J.~S.}\ \bibnamefont
  {Clarke}}, \ and\ \bibinfo {author} {\bibfnamefont {M.}~\bibnamefont
  {Veldhorst}},\ }\href@noop {} {\bibfield  {journal} {\bibinfo  {journal}
  {Science Advances}\ }\textbf {\bibinfo {volume} {4}},\ \bibinfo {pages}
  {eaar3960} (\bibinfo {year} {2018})}\BibitemShut {NoStop}%
\bibitem [{\citenamefont {Xue}\ \emph {et~al.}(2022)\citenamefont {Xue},
  \citenamefont {Russ}, \citenamefont {Samkharadze}, \citenamefont {Undseth},
  \citenamefont {Sammak}, \citenamefont {Scappucci},\ and\ \citenamefont
  {Vandersypen}}]{xue2022quantum}%
  \BibitemOpen
  \bibfield  {author} {\bibinfo {author} {\bibfnamefont {X.}~\bibnamefont
  {Xue}}, \bibinfo {author} {\bibfnamefont {M.}~\bibnamefont {Russ}}, \bibinfo
  {author} {\bibfnamefont {N.}~\bibnamefont {Samkharadze}}, \bibinfo {author}
  {\bibfnamefont {B.}~\bibnamefont {Undseth}}, \bibinfo {author} {\bibfnamefont
  {A.}~\bibnamefont {Sammak}}, \bibinfo {author} {\bibfnamefont
  {G.}~\bibnamefont {Scappucci}}, \ and\ \bibinfo {author} {\bibfnamefont
  {L.~M.}\ \bibnamefont {Vandersypen}},\ }\href@noop {} {\bibfield  {journal}
  {\bibinfo  {journal} {Nature}\ }\textbf {\bibinfo {volume} {601}},\ \bibinfo
  {pages} {343} (\bibinfo {year} {2022})}\BibitemShut {NoStop}%
\bibitem [{\citenamefont {Noiri}\ \emph {et~al.}(2022)\citenamefont {Noiri},
  \citenamefont {Takeda}, \citenamefont {Nakajima}, \citenamefont {Kobayashi},
  \citenamefont {Sammak}, \citenamefont {Scappucci},\ and\ \citenamefont
  {Tarucha}}]{noiri2022fast}%
  \BibitemOpen
  \bibfield  {author} {\bibinfo {author} {\bibfnamefont {A.}~\bibnamefont
  {Noiri}}, \bibinfo {author} {\bibfnamefont {K.}~\bibnamefont {Takeda}},
  \bibinfo {author} {\bibfnamefont {T.}~\bibnamefont {Nakajima}}, \bibinfo
  {author} {\bibfnamefont {T.}~\bibnamefont {Kobayashi}}, \bibinfo {author}
  {\bibfnamefont {A.}~\bibnamefont {Sammak}}, \bibinfo {author} {\bibfnamefont
  {G.}~\bibnamefont {Scappucci}}, \ and\ \bibinfo {author} {\bibfnamefont
  {S.}~\bibnamefont {Tarucha}},\ }\href@noop {} {\bibfield  {journal} {\bibinfo
   {journal} {Nature}\ }\textbf {\bibinfo {volume} {601}},\ \bibinfo {pages}
  {338} (\bibinfo {year} {2022})}\BibitemShut {NoStop}%
\bibitem [{\citenamefont {Mills}\ \emph {et~al.}(2021)\citenamefont {Mills},
  \citenamefont {Guinn}, \citenamefont {Gullans}, \citenamefont {Sigillito},
  \citenamefont {Feldman}, \citenamefont {Nielsen},\ and\ \citenamefont
  {Petta}}]{mills2021two}%
  \BibitemOpen
  \bibfield  {author} {\bibinfo {author} {\bibfnamefont {A.}~\bibnamefont
  {Mills}}, \bibinfo {author} {\bibfnamefont {C.}~\bibnamefont {Guinn}},
  \bibinfo {author} {\bibfnamefont {M.}~\bibnamefont {Gullans}}, \bibinfo
  {author} {\bibfnamefont {A.}~\bibnamefont {Sigillito}}, \bibinfo {author}
  {\bibfnamefont {M.}~\bibnamefont {Feldman}}, \bibinfo {author} {\bibfnamefont
  {E.}~\bibnamefont {Nielsen}}, \ and\ \bibinfo {author} {\bibfnamefont
  {J.}~\bibnamefont {Petta}},\ }\href@noop {} {\bibfield  {journal} {\bibinfo
  {journal} {arXiv preprint arXiv:2111.11937}\ } (\bibinfo {year}
  {2021})}\BibitemShut {NoStop}%
\bibitem [{\citenamefont {Harvey-Collard}\ \emph {et~al.}(2021)\citenamefont
  {Harvey-Collard}, \citenamefont {Dijkema}, \citenamefont {Zheng},
  \citenamefont {Sammak}, \citenamefont {Scappucci},\ and\ \citenamefont
  {Vandersypen}}]{harvey-collard_circuit_2021}%
  \BibitemOpen
  \bibfield  {author} {\bibinfo {author} {\bibfnamefont {P.}~\bibnamefont
  {Harvey-Collard}}, \bibinfo {author} {\bibfnamefont {J.}~\bibnamefont
  {Dijkema}}, \bibinfo {author} {\bibfnamefont {G.}~\bibnamefont {Zheng}},
  \bibinfo {author} {\bibfnamefont {A.}~\bibnamefont {Sammak}}, \bibinfo
  {author} {\bibfnamefont {G.}~\bibnamefont {Scappucci}}, \ and\ \bibinfo
  {author} {\bibfnamefont {L.~M.~K.}\ \bibnamefont {Vandersypen}},\ }\href@noop
  {} {\bibfield  {journal} {\bibinfo  {journal} {Preprint at
  http://arxiv.org/abs/2108.01206}\ } (\bibinfo {year} {2021})}\BibitemShut
  {NoStop}%
\bibitem [{\citenamefont {Xue}\ \emph {et~al.}(2021)\citenamefont {Xue},
  \citenamefont {Patra}, \citenamefont {van Dijk}, \citenamefont {Samkharadze},
  \citenamefont {Subramanian}, \citenamefont {Corna}, \citenamefont {Wuetz},
  \citenamefont {Jeon}, \citenamefont {Sheikh}, \citenamefont
  {Juarez-Hernandez} \emph {et~al.}}]{xue2021cmos}%
  \BibitemOpen
  \bibfield  {author} {\bibinfo {author} {\bibfnamefont {X.}~\bibnamefont
  {Xue}}, \bibinfo {author} {\bibfnamefont {B.}~\bibnamefont {Patra}}, \bibinfo
  {author} {\bibfnamefont {J.~P.}\ \bibnamefont {van Dijk}}, \bibinfo {author}
  {\bibfnamefont {N.}~\bibnamefont {Samkharadze}}, \bibinfo {author}
  {\bibfnamefont {S.}~\bibnamefont {Subramanian}}, \bibinfo {author}
  {\bibfnamefont {A.}~\bibnamefont {Corna}}, \bibinfo {author} {\bibfnamefont
  {B.~P.}\ \bibnamefont {Wuetz}}, \bibinfo {author} {\bibfnamefont
  {C.}~\bibnamefont {Jeon}}, \bibinfo {author} {\bibfnamefont {F.}~\bibnamefont
  {Sheikh}}, \bibinfo {author} {\bibfnamefont {E.}~\bibnamefont
  {Juarez-Hernandez}},  \emph {et~al.},\ }\href@noop {} {\bibfield  {journal}
  {\bibinfo  {journal} {Nature}\ }\textbf {\bibinfo {volume} {593}},\ \bibinfo
  {pages} {205} (\bibinfo {year} {2021})}\BibitemShut {NoStop}%
\bibitem [{\citenamefont {Lawrie}\ \emph {et~al.}(2020)\citenamefont {Lawrie},
  \citenamefont {Eenink}, \citenamefont {Hendrickx}, \citenamefont {Boter},
  \citenamefont {Petit}, \citenamefont {Amitonov}, \citenamefont {Lodari},
  \citenamefont {Paquelet~Wuetz}, \citenamefont {Volk}, \citenamefont {Philips}
  \emph {et~al.}}]{lawrie2020quantum}%
  \BibitemOpen
  \bibfield  {author} {\bibinfo {author} {\bibfnamefont {W.}~\bibnamefont
  {Lawrie}}, \bibinfo {author} {\bibfnamefont {H.}~\bibnamefont {Eenink}},
  \bibinfo {author} {\bibfnamefont {N.}~\bibnamefont {Hendrickx}}, \bibinfo
  {author} {\bibfnamefont {J.}~\bibnamefont {Boter}}, \bibinfo {author}
  {\bibfnamefont {L.}~\bibnamefont {Petit}}, \bibinfo {author} {\bibfnamefont
  {S.}~\bibnamefont {Amitonov}}, \bibinfo {author} {\bibfnamefont
  {M.}~\bibnamefont {Lodari}}, \bibinfo {author} {\bibfnamefont
  {B.}~\bibnamefont {Paquelet~Wuetz}}, \bibinfo {author} {\bibfnamefont
  {C.}~\bibnamefont {Volk}}, \bibinfo {author} {\bibfnamefont {S.}~\bibnamefont
  {Philips}},  \emph {et~al.},\ }\href@noop {} {\bibfield  {journal} {\bibinfo
  {journal} {Applied Physics Letters}\ }\textbf {\bibinfo {volume} {116}},\
  \bibinfo {pages} {080501} (\bibinfo {year} {2020})}\BibitemShut {NoStop}%
\bibitem [{\citenamefont {Thorbeck}\ and\ \citenamefont
  {Zimmerman}(2015)}]{thorbeck2015formation}%
  \BibitemOpen
  \bibfield  {author} {\bibinfo {author} {\bibfnamefont {T.}~\bibnamefont
  {Thorbeck}}\ and\ \bibinfo {author} {\bibfnamefont {N.~M.}\ \bibnamefont
  {Zimmerman}},\ }\href@noop {} {\bibfield  {journal} {\bibinfo  {journal} {AIP
  Advances}\ }\textbf {\bibinfo {volume} {5}},\ \bibinfo {pages} {087107}
  (\bibinfo {year} {2015})}\BibitemShut {NoStop}%
\bibitem [{\citenamefont {Connors}\ \emph {et~al.}(2019)\citenamefont
  {Connors}, \citenamefont {Nelson}, \citenamefont {Qiao}, \citenamefont
  {Edge},\ and\ \citenamefont {Nichol}}]{connors2019low}%
  \BibitemOpen
  \bibfield  {author} {\bibinfo {author} {\bibfnamefont {E.~J.}\ \bibnamefont
  {Connors}}, \bibinfo {author} {\bibfnamefont {J.}~\bibnamefont {Nelson}},
  \bibinfo {author} {\bibfnamefont {H.}~\bibnamefont {Qiao}}, \bibinfo {author}
  {\bibfnamefont {L.~F.}\ \bibnamefont {Edge}}, \ and\ \bibinfo {author}
  {\bibfnamefont {J.~M.}\ \bibnamefont {Nichol}},\ }\href@noop {} {\bibfield
  {journal} {\bibinfo  {journal} {Physical Review B}\ }\textbf {\bibinfo
  {volume} {100}},\ \bibinfo {pages} {165305} (\bibinfo {year}
  {2019})}\BibitemShut {NoStop}%
\bibitem [{\citenamefont {Struck}\ \emph {et~al.}(2020)\citenamefont {Struck},
  \citenamefont {Hollmann}, \citenamefont {Schauer}, \citenamefont {Fedorets},
  \citenamefont {Schmidbauer}, \citenamefont {Sawano}, \citenamefont {Riemann},
  \citenamefont {Abrosimov}, \citenamefont {Cywinski}, \citenamefont
  {Bougeard},\ and\ \citenamefont {Schreiber}}]{struck_lowfrequency_2020}%
  \BibitemOpen
  \bibfield  {author} {\bibinfo {author} {\bibfnamefont {T.}~\bibnamefont
  {Struck}}, \bibinfo {author} {\bibfnamefont {A.}~\bibnamefont {Hollmann}},
  \bibinfo {author} {\bibfnamefont {F.}~\bibnamefont {Schauer}}, \bibinfo
  {author} {\bibfnamefont {O.}~\bibnamefont {Fedorets}}, \bibinfo {author}
  {\bibfnamefont {A.}~\bibnamefont {Schmidbauer}}, \bibinfo {author}
  {\bibfnamefont {K.}~\bibnamefont {Sawano}}, \bibinfo {author} {\bibfnamefont
  {H.}~\bibnamefont {Riemann}}, \bibinfo {author} {\bibfnamefont {N.~V.}\
  \bibnamefont {Abrosimov}}, \bibinfo {author} {\bibfnamefont {L.}~\bibnamefont
  {Cywinski}}, \bibinfo {author} {\bibfnamefont {D.}~\bibnamefont {Bougeard}},
  \ and\ \bibinfo {author} {\bibfnamefont {L.~R.}\ \bibnamefont {Schreiber}},\
  }\href@noop {} {\bibfield  {journal} {\bibinfo  {journal} {npj Quantum
  Information}\ }\textbf {\bibinfo {volume} {6}},\ \bibinfo {pages} {40}
  (\bibinfo {year} {2020})}\BibitemShut {NoStop}%
\bibitem [{\citenamefont {Paquelet~Wuetz}\ \emph {et~al.}(2021)\citenamefont
  {Paquelet~Wuetz}, \citenamefont {Losert}, \citenamefont {Koelling},
  \citenamefont {Stehouwer}, \citenamefont {Zwerver}, \citenamefont {Philips},
  \citenamefont {Madzik}, \citenamefont {Xue}, \citenamefont {Zheng},
  \citenamefont {Lodari} \emph {et~al.}}]{wuetz2021atomic}%
  \BibitemOpen
  \bibfield  {author} {\bibinfo {author} {\bibfnamefont {B.}~\bibnamefont
  {Paquelet~Wuetz}}, \bibinfo {author} {\bibfnamefont {M.~P.}\ \bibnamefont
  {Losert}}, \bibinfo {author} {\bibfnamefont {S.}~\bibnamefont {Koelling}},
  \bibinfo {author} {\bibfnamefont {L.~E.}\ \bibnamefont {Stehouwer}}, \bibinfo
  {author} {\bibfnamefont {A.-M.~J.}\ \bibnamefont {Zwerver}}, \bibinfo
  {author} {\bibfnamefont {S.~G.}\ \bibnamefont {Philips}}, \bibinfo {author}
  {\bibfnamefont {M.~T.}\ \bibnamefont {Madzik}}, \bibinfo {author}
  {\bibfnamefont {X.}~\bibnamefont {Xue}}, \bibinfo {author} {\bibfnamefont
  {G.}~\bibnamefont {Zheng}}, \bibinfo {author} {\bibfnamefont
  {M.}~\bibnamefont {Lodari}},  \emph {et~al.},\ }\href@noop {} {\bibfield
  {journal} {\bibinfo  {journal} {arXiv preprint arXiv:2112.09606}\ } (\bibinfo
  {year} {2021})}\BibitemShut {NoStop}%
\bibitem [{\citenamefont {Chen}\ \emph {et~al.}(2021)\citenamefont {Chen},
  \citenamefont {Raach}, \citenamefont {Pan}, \citenamefont {Kiselev},
  \citenamefont {Acuna}, \citenamefont {Blumoff}, \citenamefont {Brecht},
  \citenamefont {Choi}, \citenamefont {Ha}, \citenamefont {Hulbert},
  \citenamefont {Jura}, \citenamefont {Keating}, \citenamefont {Noah},
  \citenamefont {Sun}, \citenamefont {Thomas}, \citenamefont {Borselli},
  \citenamefont {Jackson}, \citenamefont {Rakher},\ and\ \citenamefont
  {Ross}}]{chen_detuning_2021}%
  \BibitemOpen
  \bibfield  {author} {\bibinfo {author} {\bibfnamefont {E.~H.}\ \bibnamefont
  {Chen}}, \bibinfo {author} {\bibfnamefont {K.}~\bibnamefont {Raach}},
  \bibinfo {author} {\bibfnamefont {A.}~\bibnamefont {Pan}}, \bibinfo {author}
  {\bibfnamefont {A.~A.}\ \bibnamefont {Kiselev}}, \bibinfo {author}
  {\bibfnamefont {E.}~\bibnamefont {Acuna}}, \bibinfo {author} {\bibfnamefont
  {J.~Z.}\ \bibnamefont {Blumoff}}, \bibinfo {author} {\bibfnamefont
  {T.}~\bibnamefont {Brecht}}, \bibinfo {author} {\bibfnamefont {M.~D.}\
  \bibnamefont {Choi}}, \bibinfo {author} {\bibfnamefont {W.}~\bibnamefont
  {Ha}}, \bibinfo {author} {\bibfnamefont {D.~R.}\ \bibnamefont {Hulbert}},
  \bibinfo {author} {\bibfnamefont {M.~P.}\ \bibnamefont {Jura}}, \bibinfo
  {author} {\bibfnamefont {T.~E.}\ \bibnamefont {Keating}}, \bibinfo {author}
  {\bibfnamefont {R.}~\bibnamefont {Noah}}, \bibinfo {author} {\bibfnamefont
  {B.}~\bibnamefont {Sun}}, \bibinfo {author} {\bibfnamefont {B.~J.}\
  \bibnamefont {Thomas}}, \bibinfo {author} {\bibfnamefont {M.~G.}\
  \bibnamefont {Borselli}}, \bibinfo {author} {\bibfnamefont {C.}~\bibnamefont
  {Jackson}}, \bibinfo {author} {\bibfnamefont {M.~T.}\ \bibnamefont {Rakher}},
  \ and\ \bibinfo {author} {\bibfnamefont {R.~S.}\ \bibnamefont {Ross}},\
  }\href {\doibase 10.1103/PhysRevApplied.15.044033} {\bibfield  {journal}
  {\bibinfo  {journal} {Physical Review Applied}\ }\textbf {\bibinfo {volume}
  {15}},\ \bibinfo {pages} {044033} (\bibinfo {year} {2021})}\BibitemShut
  {NoStop}%
\bibitem [{\citenamefont {Hollmann}\ \emph {et~al.}(2020)\citenamefont
  {Hollmann}, \citenamefont {Struck}, \citenamefont {Langrock}, \citenamefont
  {Schmidbauer}, \citenamefont {Schauer}, \citenamefont {Leonhardt},
  \citenamefont {Sawano}, \citenamefont {Riemann}, \citenamefont {Abrosimov},
  \citenamefont {Bougeard},\ and\ \citenamefont
  {Schreiber}}]{hollmann_large_2020}%
  \BibitemOpen
  \bibfield  {author} {\bibinfo {author} {\bibfnamefont {A.}~\bibnamefont
  {Hollmann}}, \bibinfo {author} {\bibfnamefont {T.}~\bibnamefont {Struck}},
  \bibinfo {author} {\bibfnamefont {V.}~\bibnamefont {Langrock}}, \bibinfo
  {author} {\bibfnamefont {A.}~\bibnamefont {Schmidbauer}}, \bibinfo {author}
  {\bibfnamefont {F.}~\bibnamefont {Schauer}}, \bibinfo {author} {\bibfnamefont
  {T.}~\bibnamefont {Leonhardt}}, \bibinfo {author} {\bibfnamefont
  {K.}~\bibnamefont {Sawano}}, \bibinfo {author} {\bibfnamefont
  {H.}~\bibnamefont {Riemann}}, \bibinfo {author} {\bibfnamefont {N.~V.}\
  \bibnamefont {Abrosimov}}, \bibinfo {author} {\bibfnamefont {D.}~\bibnamefont
  {Bougeard}}, \ and\ \bibinfo {author} {\bibfnamefont {L.~R.}\ \bibnamefont
  {Schreiber}},\ }\href {\doibase 10.1103/PhysRevApplied.13.034068} {\bibfield
  {journal} {\bibinfo  {journal} {Physical Review Applied}\ }\textbf {\bibinfo
  {volume} {13}},\ \bibinfo {pages} {034068} (\bibinfo {year}
  {2020})}\BibitemShut {NoStop}%
\bibitem [{\citenamefont {Sabbagh}\ \emph {et~al.}(2019)\citenamefont
  {Sabbagh}, \citenamefont {Thomas}, \citenamefont {Torres}, \citenamefont
  {Pillarisetty}, \citenamefont {Amin}, \citenamefont {George}, \citenamefont
  {Singh}, \citenamefont {Budrevich}, \citenamefont {Robinson}, \citenamefont
  {Merrill}, \citenamefont {Ross}, \citenamefont {Roberts}, \citenamefont
  {Lampert}, \citenamefont {Massa}, \citenamefont {Amitonov}, \citenamefont
  {Boter}, \citenamefont {Droulers}, \citenamefont {Eenink}, \citenamefont {van
  Hezel}, \citenamefont {Donelson}, \citenamefont {Veldhorst}, \citenamefont
  {Vandersypen}, \citenamefont {Clarke},\ and\ \citenamefont
  {Scappucci}}]{SabbaghPhysRevApplied2019}%
  \BibitemOpen
  \bibfield  {author} {\bibinfo {author} {\bibfnamefont {D.}~\bibnamefont
  {Sabbagh}}, \bibinfo {author} {\bibfnamefont {N.}~\bibnamefont {Thomas}},
  \bibinfo {author} {\bibfnamefont {J.}~\bibnamefont {Torres}}, \bibinfo
  {author} {\bibfnamefont {R.}~\bibnamefont {Pillarisetty}}, \bibinfo {author}
  {\bibfnamefont {P.}~\bibnamefont {Amin}}, \bibinfo {author} {\bibfnamefont
  {H.}~\bibnamefont {George}}, \bibinfo {author} {\bibfnamefont
  {K.}~\bibnamefont {Singh}}, \bibinfo {author} {\bibfnamefont
  {A.}~\bibnamefont {Budrevich}}, \bibinfo {author} {\bibfnamefont
  {M.}~\bibnamefont {Robinson}}, \bibinfo {author} {\bibfnamefont
  {D.}~\bibnamefont {Merrill}}, \bibinfo {author} {\bibfnamefont
  {L.}~\bibnamefont {Ross}}, \bibinfo {author} {\bibfnamefont {J.}~\bibnamefont
  {Roberts}}, \bibinfo {author} {\bibfnamefont {L.}~\bibnamefont {Lampert}},
  \bibinfo {author} {\bibfnamefont {L.}~\bibnamefont {Massa}}, \bibinfo
  {author} {\bibfnamefont {S.}~\bibnamefont {Amitonov}}, \bibinfo {author}
  {\bibfnamefont {J.}~\bibnamefont {Boter}}, \bibinfo {author} {\bibfnamefont
  {G.}~\bibnamefont {Droulers}}, \bibinfo {author} {\bibfnamefont
  {H.}~\bibnamefont {Eenink}}, \bibinfo {author} {\bibfnamefont
  {M.}~\bibnamefont {van Hezel}}, \bibinfo {author} {\bibfnamefont
  {D.}~\bibnamefont {Donelson}}, \bibinfo {author} {\bibfnamefont
  {M.}~\bibnamefont {Veldhorst}}, \bibinfo {author} {\bibfnamefont
  {L.}~\bibnamefont {Vandersypen}}, \bibinfo {author} {\bibfnamefont
  {J.}~\bibnamefont {Clarke}}, \ and\ \bibinfo {author} {\bibfnamefont
  {G.}~\bibnamefont {Scappucci}},\ }\href {\doibase
  10.1103/PhysRevApplied.12.014013} {\bibfield  {journal} {\bibinfo  {journal}
  {Phys. Rev. Applied}\ }\textbf {\bibinfo {volume} {12}},\ \bibinfo {pages}
  {014013} (\bibinfo {year} {2019})}\BibitemShut {NoStop}%
\bibitem [{Note1()}]{Note1}%
  \BibitemOpen
  \bibinfo {note} {A typical secondary ions mass spectrometry of our
  heterostructures is reported in Fig.~S13 of Ref.~\cite {wuetz2021atomic}. The
  oxygen concentration in the $^{28}$Si quantum well is $\simeq 4 \times
  10^{17}$~cm$^{-3}$}\BibitemShut {NoStop}%
\bibitem [{\citenamefont {Samkharadze}\ \emph {et~al.}(2018)\citenamefont
  {Samkharadze}, \citenamefont {Zheng}, \citenamefont {Kalhor}, \citenamefont
  {Brousse}, \citenamefont {Sammak}, \citenamefont {Mendes}, \citenamefont
  {Blais}, \citenamefont {Scappucci},\ and\ \citenamefont
  {Vandersypen}}]{samkharadze2018strong}%
  \BibitemOpen
  \bibfield  {author} {\bibinfo {author} {\bibfnamefont {N.}~\bibnamefont
  {Samkharadze}}, \bibinfo {author} {\bibfnamefont {G.}~\bibnamefont {Zheng}},
  \bibinfo {author} {\bibfnamefont {N.}~\bibnamefont {Kalhor}}, \bibinfo
  {author} {\bibfnamefont {D.}~\bibnamefont {Brousse}}, \bibinfo {author}
  {\bibfnamefont {A.}~\bibnamefont {Sammak}}, \bibinfo {author} {\bibfnamefont
  {U.}~\bibnamefont {Mendes}}, \bibinfo {author} {\bibfnamefont
  {A.}~\bibnamefont {Blais}}, \bibinfo {author} {\bibfnamefont
  {G.}~\bibnamefont {Scappucci}}, \ and\ \bibinfo {author} {\bibfnamefont
  {L.}~\bibnamefont {Vandersypen}},\ }\href@noop {} {\bibfield  {journal}
  {\bibinfo  {journal} {Science}\ }\textbf {\bibinfo {volume} {359}},\ \bibinfo
  {pages} {1123} (\bibinfo {year} {2018})}\BibitemShut {NoStop}%
\bibitem [{\citenamefont {Gao}\ \emph {et~al.}(1993)\citenamefont {Gao},
  \citenamefont {Cheng}, \citenamefont {Chen}, \citenamefont {Choyke},\ and\
  \citenamefont {Yates}}]{gao_chlorine_1993}%
  \BibitemOpen
  \bibfield  {author} {\bibinfo {author} {\bibfnamefont {Q.}~\bibnamefont
  {Gao}}, \bibinfo {author} {\bibfnamefont {C.~C.}\ \bibnamefont {Cheng}},
  \bibinfo {author} {\bibfnamefont {P.~J.}\ \bibnamefont {Chen}}, \bibinfo
  {author} {\bibfnamefont {W.~J.}\ \bibnamefont {Choyke}}, \ and\ \bibinfo
  {author} {\bibfnamefont {J.~T.}\ \bibnamefont {Yates}},\ }\href {\doibase
  10.1063/1.464536} {\bibfield  {journal} {\bibinfo  {journal} {The Journal of
  Chemical Physics}\ }\textbf {\bibinfo {volume} {98}},\ \bibinfo {pages}
  {8308} (\bibinfo {year} {1993})}\BibitemShut {NoStop}%
\bibitem [{\citenamefont {Bolotov}\ \emph {et~al.}(2021)\citenamefont
  {Bolotov}, \citenamefont {Fons}, \citenamefont {Mimura}, \citenamefont
  {Sasaki},\ and\ \citenamefont {Uchida}}]{bolotov_electronic_2021}%
  \BibitemOpen
  \bibfield  {author} {\bibinfo {author} {\bibfnamefont {L.}~\bibnamefont
  {Bolotov}}, \bibinfo {author} {\bibfnamefont {P.}~\bibnamefont {Fons}},
  \bibinfo {author} {\bibfnamefont {H.}~\bibnamefont {Mimura}}, \bibinfo
  {author} {\bibfnamefont {T.}~\bibnamefont {Sasaki}}, \ and\ \bibinfo {author}
  {\bibfnamefont {N.}~\bibnamefont {Uchida}},\ }\href {\doibase
  10.1016/j.apsusc.2021.151135} {\bibfield  {journal} {\bibinfo  {journal}
  {Applied Surface Science}\ }\textbf {\bibinfo {volume} {570}},\ \bibinfo
  {pages} {151135} (\bibinfo {year} {2021})}\BibitemShut {NoStop}%
\bibitem [{\citenamefont {Hartmann}\ \emph {et~al.}(2008)\citenamefont
  {Hartmann}, \citenamefont {Andrieu}, \citenamefont {Lafond}, \citenamefont
  {Ernst}, \citenamefont {Bogumilowicz}, \citenamefont {Delaye}, \citenamefont
  {Weber}, \citenamefont {Rouchon}, \citenamefont {Papon},\ and\ \citenamefont
  {Cherkashin}}]{hartmann_reduced_2008}%
  \BibitemOpen
  \bibfield  {author} {\bibinfo {author} {\bibfnamefont {J.~M.}\ \bibnamefont
  {Hartmann}}, \bibinfo {author} {\bibfnamefont {F.}~\bibnamefont {Andrieu}},
  \bibinfo {author} {\bibfnamefont {D.}~\bibnamefont {Lafond}}, \bibinfo
  {author} {\bibfnamefont {T.}~\bibnamefont {Ernst}}, \bibinfo {author}
  {\bibfnamefont {Y.}~\bibnamefont {Bogumilowicz}}, \bibinfo {author}
  {\bibfnamefont {V.}~\bibnamefont {Delaye}}, \bibinfo {author} {\bibfnamefont
  {O.}~\bibnamefont {Weber}}, \bibinfo {author} {\bibfnamefont
  {D.}~\bibnamefont {Rouchon}}, \bibinfo {author} {\bibfnamefont {A.~M.}\
  \bibnamefont {Papon}}, \ and\ \bibinfo {author} {\bibfnamefont
  {N.}~\bibnamefont {Cherkashin}},\ }\href {\doibase
  10.1016/j.mseb.2008.08.009} {\bibfield  {journal} {\bibinfo  {journal}
  {Materials Science and Engineering: B}\ }\bibinfo {series} {Front-{End}
  {Junction} and {Contact} {Formation} in {Future} {Silicon}/{Germanium}
  {Based} {Devices}},\ \textbf {\bibinfo {volume} {154-155}},\ \bibinfo {pages}
  {76} (\bibinfo {year} {2008})}\BibitemShut {NoStop}%
\bibitem [{\citenamefont {Hartmann}\ \emph {et~al.}(2009)\citenamefont
  {Hartmann}, \citenamefont {Abbadie}, \citenamefont {Cherkashin},
  \citenamefont {Grampeix},\ and\ \citenamefont
  {Clavelier}}]{hartmann_epitaxial_2009}%
  \BibitemOpen
  \bibfield  {author} {\bibinfo {author} {\bibfnamefont {J.~M.}\ \bibnamefont
  {Hartmann}}, \bibinfo {author} {\bibfnamefont {A.}~\bibnamefont {Abbadie}},
  \bibinfo {author} {\bibfnamefont {N.}~\bibnamefont {Cherkashin}}, \bibinfo
  {author} {\bibfnamefont {H.}~\bibnamefont {Grampeix}}, \ and\ \bibinfo
  {author} {\bibfnamefont {L.}~\bibnamefont {Clavelier}},\ }\href {\doibase
  10.1088/0268-1242/24/5/055002} {\bibfield  {journal} {\bibinfo  {journal}
  {Semiconductor Science and Technology}\ }\textbf {\bibinfo {volume} {24}},\
  \bibinfo {pages} {055002} (\bibinfo {year} {2009})}\BibitemShut {NoStop}%
\bibitem [{\citenamefont {Bauer}\ and\ \citenamefont
  {Thomas}(2010)}]{bauer_novel_2010}%
  \BibitemOpen
  \bibfield  {author} {\bibinfo {author} {\bibfnamefont {M.}~\bibnamefont
  {Bauer}}\ and\ \bibinfo {author} {\bibfnamefont {S.}~\bibnamefont {Thomas}},\
  }\href {\doibase 10.1016/j.tsf.2009.10.088} {\bibfield  {journal} {\bibinfo
  {journal} {Thin Solid Films}\ }\textbf {\bibinfo {volume} {518}},\ \bibinfo
  {pages} {S200} (\bibinfo {year} {2010})}\BibitemShut {NoStop}%
\bibitem [{\citenamefont {Vincent}\ \emph {et~al.}(2010)\citenamefont
  {Vincent}, \citenamefont {Loo}, \citenamefont {Vandervorst}, \citenamefont
  {Brammertz},\ and\ \citenamefont {Caymax}}]{vincent_low_2010}%
  \BibitemOpen
  \bibfield  {author} {\bibinfo {author} {\bibfnamefont {B.}~\bibnamefont
  {Vincent}}, \bibinfo {author} {\bibfnamefont {R.}~\bibnamefont {Loo}},
  \bibinfo {author} {\bibfnamefont {W.}~\bibnamefont {Vandervorst}}, \bibinfo
  {author} {\bibfnamefont {G.}~\bibnamefont {Brammertz}}, \ and\ \bibinfo
  {author} {\bibfnamefont {M.}~\bibnamefont {Caymax}},\ }\href {\doibase
  10.1016/j.jcrysgro.2010.06.013} {\bibfield  {journal} {\bibinfo  {journal}
  {Journal of Crystal Growth}\ }\textbf {\bibinfo {volume} {312}},\ \bibinfo
  {pages} {2671} (\bibinfo {year} {2010})}\BibitemShut {NoStop}%
\bibitem [{\citenamefont {Vincent}\ \emph {et~al.}(2011)\citenamefont
  {Vincent}, \citenamefont {Loo}, \citenamefont {Vandervorst}, \citenamefont
  {Delmotte}, \citenamefont {Douhard}, \citenamefont {Valev}, \citenamefont
  {Vanbel}, \citenamefont {Verbiest}, \citenamefont {Rip}, \citenamefont
  {Brijs}, \citenamefont {Conard}, \citenamefont {Claypool}, \citenamefont
  {Takeuchi}, \citenamefont {Zaima}, \citenamefont {Mitard}, \citenamefont
  {De~Jaeger}, \citenamefont {Dekoster},\ and\ \citenamefont
  {Caymax}}]{vincent_si_2011}%
  \BibitemOpen
  \bibfield  {author} {\bibinfo {author} {\bibfnamefont {B.}~\bibnamefont
  {Vincent}}, \bibinfo {author} {\bibfnamefont {R.}~\bibnamefont {Loo}},
  \bibinfo {author} {\bibfnamefont {W.}~\bibnamefont {Vandervorst}}, \bibinfo
  {author} {\bibfnamefont {J.}~\bibnamefont {Delmotte}}, \bibinfo {author}
  {\bibfnamefont {B.}~\bibnamefont {Douhard}}, \bibinfo {author} {\bibfnamefont
  {V.}~\bibnamefont {Valev}}, \bibinfo {author} {\bibfnamefont
  {M.}~\bibnamefont {Vanbel}}, \bibinfo {author} {\bibfnamefont
  {T.}~\bibnamefont {Verbiest}}, \bibinfo {author} {\bibfnamefont
  {J.}~\bibnamefont {Rip}}, \bibinfo {author} {\bibfnamefont {B.}~\bibnamefont
  {Brijs}}, \bibinfo {author} {\bibfnamefont {T.}~\bibnamefont {Conard}},
  \bibinfo {author} {\bibfnamefont {C.}~\bibnamefont {Claypool}}, \bibinfo
  {author} {\bibfnamefont {S.}~\bibnamefont {Takeuchi}}, \bibinfo {author}
  {\bibfnamefont {S.}~\bibnamefont {Zaima}}, \bibinfo {author} {\bibfnamefont
  {J.}~\bibnamefont {Mitard}}, \bibinfo {author} {\bibfnamefont
  {B.}~\bibnamefont {De~Jaeger}}, \bibinfo {author} {\bibfnamefont
  {J.}~\bibnamefont {Dekoster}}, \ and\ \bibinfo {author} {\bibfnamefont
  {M.}~\bibnamefont {Caymax}},\ }\href {\doibase 10.1016/j.sse.2011.01.049}
  {\bibfield  {journal} {\bibinfo  {journal} {Solid-State Electronics}\
  }\textbf {\bibinfo {volume} {60}},\ \bibinfo {pages} {116} (\bibinfo {year}
  {2011})}\BibitemShut {NoStop}%
\bibitem [{\citenamefont {Hartmann}\ \emph {et~al.}(2012)\citenamefont
  {Hartmann}, \citenamefont {Benevent}, \citenamefont {Damlencourt},\ and\
  \citenamefont {Billon}}]{hartmann_benchmarking_2012}%
  \BibitemOpen
  \bibfield  {author} {\bibinfo {author} {\bibfnamefont {J.~M.}\ \bibnamefont
  {Hartmann}}, \bibinfo {author} {\bibfnamefont {V.}~\bibnamefont {Benevent}},
  \bibinfo {author} {\bibfnamefont {J.~F.}\ \bibnamefont {Damlencourt}}, \ and\
  \bibinfo {author} {\bibfnamefont {T.}~\bibnamefont {Billon}},\ }\href
  {\doibase 10.1016/j.tsf.2011.10.164} {\bibfield  {journal} {\bibinfo
  {journal} {Thin Solid Films}\ }\bibinfo {series} {{ICSI}-7},\ \textbf
  {\bibinfo {volume} {520}},\ \bibinfo {pages} {3185} (\bibinfo {year}
  {2012})}\BibitemShut {NoStop}%
\bibitem [{\citenamefont {Aubin}\ \emph {et~al.}(2016)\citenamefont {Aubin},
  \citenamefont {Hartmann},\ and\ \citenamefont
  {Benevent}}]{aubin_epitaxial_2016}%
  \BibitemOpen
  \bibfield  {author} {\bibinfo {author} {\bibfnamefont {J.}~\bibnamefont
  {Aubin}}, \bibinfo {author} {\bibfnamefont {J.~M.}\ \bibnamefont {Hartmann}},
  \ and\ \bibinfo {author} {\bibfnamefont {V.}~\bibnamefont {Benevent}},\
  }\href {\doibase 10.1016/j.tsf.2015.07.024} {\bibfield  {journal} {\bibinfo
  {journal} {Thin Solid Films}\ }\bibinfo {series} {The 9th {International}
  {Conference} on {Silicon} {Epitaxy} and {Heterostructures}},\ \textbf
  {\bibinfo {volume} {602}},\ \bibinfo {pages} {36} (\bibinfo {year}
  {2016})}\BibitemShut {NoStop}%
\bibitem [{\citenamefont {Liu}\ \emph {et~al.}(1995)\citenamefont {Liu},
  \citenamefont {Nicolet}, \citenamefont {Park}, \citenamefont {Koak},\ and\
  \citenamefont {Lee}}]{liu1995hydridation}%
  \BibitemOpen
  \bibfield  {author} {\bibinfo {author} {\bibfnamefont {W.}~\bibnamefont
  {Liu}}, \bibinfo {author} {\bibfnamefont {M.-A.}\ \bibnamefont {Nicolet}},
  \bibinfo {author} {\bibfnamefont {H.-H.}\ \bibnamefont {Park}}, \bibinfo
  {author} {\bibfnamefont {B.-H.}\ \bibnamefont {Koak}}, \ and\ \bibinfo
  {author} {\bibfnamefont {J.-W.}\ \bibnamefont {Lee}},\ }\href@noop {}
  {\bibfield  {journal} {\bibinfo  {journal} {Journal of Applied Physics}\
  }\textbf {\bibinfo {volume} {78}},\ \bibinfo {pages} {2631} (\bibinfo {year}
  {1995})}\BibitemShut {NoStop}%
\bibitem [{\citenamefont {Long}\ \emph {et~al.}(2012)\citenamefont {Long},
  \citenamefont {Azarov}, \citenamefont {Kløw}, \citenamefont {Galeckas},
  \citenamefont {Yu~Kuznetsov},\ and\ \citenamefont {Diplas}}]{long_ge_2012}%
  \BibitemOpen
  \bibfield  {author} {\bibinfo {author} {\bibfnamefont {E.}~\bibnamefont
  {Long}}, \bibinfo {author} {\bibfnamefont {A.}~\bibnamefont {Azarov}},
  \bibinfo {author} {\bibfnamefont {F.}~\bibnamefont {Kløw}}, \bibinfo
  {author} {\bibfnamefont {A.}~\bibnamefont {Galeckas}}, \bibinfo {author}
  {\bibfnamefont {A.}~\bibnamefont {Yu~Kuznetsov}}, \ and\ \bibinfo {author}
  {\bibfnamefont {S.}~\bibnamefont {Diplas}},\ }\href {\doibase
  10.1063/1.3677987} {\bibfield  {journal} {\bibinfo  {journal} {Journal of
  Applied Physics}\ }\textbf {\bibinfo {volume} {111}},\ \bibinfo {pages}
  {024308} (\bibinfo {year} {2012})}\BibitemShut {NoStop}%
\bibitem [{\citenamefont {Liou}\ \emph {et~al.}(1991)\citenamefont {Liou},
  \citenamefont {Mei}, \citenamefont {Gennser},\ and\ \citenamefont
  {Yang}}]{liou1991effects}%
  \BibitemOpen
  \bibfield  {author} {\bibinfo {author} {\bibfnamefont {H.}~\bibnamefont
  {Liou}}, \bibinfo {author} {\bibfnamefont {P.}~\bibnamefont {Mei}}, \bibinfo
  {author} {\bibfnamefont {U.}~\bibnamefont {Gennser}}, \ and\ \bibinfo
  {author} {\bibfnamefont {E.}~\bibnamefont {Yang}},\ }\href@noop {} {\bibfield
   {journal} {\bibinfo  {journal} {Applied physics letters}\ }\textbf {\bibinfo
  {volume} {59}},\ \bibinfo {pages} {1200} (\bibinfo {year}
  {1991})}\BibitemShut {NoStop}%
\bibitem [{\citenamefont {LeGoues}\ \emph {et~al.}(1989)\citenamefont
  {LeGoues}, \citenamefont {Rosenberg}, \citenamefont {Nguyen}, \citenamefont
  {Himpsel},\ and\ \citenamefont {Meyerson}}]{legoues1989oxidation}%
  \BibitemOpen
  \bibfield  {author} {\bibinfo {author} {\bibfnamefont {F.}~\bibnamefont
  {LeGoues}}, \bibinfo {author} {\bibfnamefont {R.}~\bibnamefont {Rosenberg}},
  \bibinfo {author} {\bibfnamefont {T.}~\bibnamefont {Nguyen}}, \bibinfo
  {author} {\bibfnamefont {F.}~\bibnamefont {Himpsel}}, \ and\ \bibinfo
  {author} {\bibfnamefont {B.}~\bibnamefont {Meyerson}},\ }\href@noop {}
  {\bibfield  {journal} {\bibinfo  {journal} {Journal of Applied Physics}\
  }\textbf {\bibinfo {volume} {65}},\ \bibinfo {pages} {1724} (\bibinfo {year}
  {1989})}\BibitemShut {NoStop}%
\bibitem [{\citenamefont {Schubert}(1994)}]{schubert1994delta}%
  \BibitemOpen
  \bibfield  {author} {\bibinfo {author} {\bibfnamefont {E.}~\bibnamefont
  {Schubert}},\ }in\ \href@noop {} {\emph {\bibinfo {booktitle} {Semiconductors
  and Semimetals}}},\ Vol.~\bibinfo {volume} {40}\ (\bibinfo  {publisher}
  {Elsevier},\ \bibinfo {year} {1994})\ pp.\ \bibinfo {pages}
  {1--151}\BibitemShut {NoStop}%
\bibitem [{Note2()}]{Note2}%
  \BibitemOpen
  \bibinfo {note} {$T = 190$~mK is the electron temperature obtained by fitting
  Coulomb blockade peaks (Supplementary) measured on quantum dot devices\cite
  {xue2021cmos} fabricated on a similar heterostructure. The electron
  temperature is higher than the temperature of 70~mK measured by a thermometer
  located on the mixing chamber of the dilution refrigerator}\BibitemShut
  {NoStop}%
\bibitem [{\citenamefont {Paquelet~Wuetz}\ \emph
  {et~al.}(2020{\natexlab{a}})\citenamefont {Paquelet~Wuetz}, \citenamefont
  {Bavdaz}, \citenamefont {Yeoh}, \citenamefont {Schouten}, \citenamefont
  {van~der Does}, \citenamefont {Tiggelman}, \citenamefont {Sabbagh},
  \citenamefont {Sammak}, \citenamefont {Almudever}, \citenamefont
  {Sebastiano}, \citenamefont {Clarke}, \citenamefont {Veldhorst},\ and\
  \citenamefont {Scappucci}}]{wuetz2019multiplexed}%
  \BibitemOpen
  \bibfield  {author} {\bibinfo {author} {\bibfnamefont {B.}~\bibnamefont
  {Paquelet~Wuetz}}, \bibinfo {author} {\bibfnamefont {P.~L.}\ \bibnamefont
  {Bavdaz}}, \bibinfo {author} {\bibfnamefont {L.~A.}\ \bibnamefont {Yeoh}},
  \bibinfo {author} {\bibfnamefont {R.}~\bibnamefont {Schouten}}, \bibinfo
  {author} {\bibfnamefont {H.}~\bibnamefont {van~der Does}}, \bibinfo {author}
  {\bibfnamefont {M.}~\bibnamefont {Tiggelman}}, \bibinfo {author}
  {\bibfnamefont {D.}~\bibnamefont {Sabbagh}}, \bibinfo {author} {\bibfnamefont
  {A.}~\bibnamefont {Sammak}}, \bibinfo {author} {\bibfnamefont {C.~G.}\
  \bibnamefont {Almudever}}, \bibinfo {author} {\bibfnamefont {F.}~\bibnamefont
  {Sebastiano}}, \bibinfo {author} {\bibfnamefont {J.~S.}\ \bibnamefont
  {Clarke}}, \bibinfo {author} {\bibfnamefont {M.}~\bibnamefont {Veldhorst}}, \
  and\ \bibinfo {author} {\bibfnamefont {G.}~\bibnamefont {Scappucci}},\ }\href
  {\doibase 10.1038/s41534-020-0274-4} {\bibfield  {journal} {\bibinfo
  {journal} {npj Quantum Information}\ }\textbf {\bibinfo {volume} {6}},\
  \bibinfo {pages} {43} (\bibinfo {year} {2020}{\natexlab{a}})}\BibitemShut
  {NoStop}%
\bibitem [{\citenamefont {Monroe}\ \emph {et~al.}(1993)\citenamefont {Monroe},
  \citenamefont {Xie}, \citenamefont {Fitzgerald}, \citenamefont {Silverman},\
  and\ \citenamefont {Watson}}]{monroe_comparison_1993}%
  \BibitemOpen
  \bibfield  {author} {\bibinfo {author} {\bibfnamefont {D.}~\bibnamefont
  {Monroe}}, \bibinfo {author} {\bibfnamefont {Y.~H.}\ \bibnamefont {Xie}},
  \bibinfo {author} {\bibfnamefont {E.~A.}\ \bibnamefont {Fitzgerald}},
  \bibinfo {author} {\bibfnamefont {P.~J.}\ \bibnamefont {Silverman}}, \ and\
  \bibinfo {author} {\bibfnamefont {G.~P.}\ \bibnamefont {Watson}},\ }\href
  {\doibase 10.1116/1.586471} {\bibfield  {journal} {\bibinfo  {journal}
  {Journal of Vacuum Science \& Technology B: Microelectronics and Nanometer
  Structures Processing, Measurement, and Phenomena}\ }\textbf {\bibinfo
  {volume} {11}},\ \bibinfo {pages} {1731} (\bibinfo {year}
  {1993})}\BibitemShut {NoStop}%
\bibitem [{\citenamefont {Mi}\ \emph {et~al.}(2015)\citenamefont {Mi},
  \citenamefont {Hazard}, \citenamefont {Payette}, \citenamefont {Wang},
  \citenamefont {Zajac}, \citenamefont {Cady},\ and\ \citenamefont
  {Petta}}]{mi2015magnetotransport}%
  \BibitemOpen
  \bibfield  {author} {\bibinfo {author} {\bibfnamefont {X.}~\bibnamefont
  {Mi}}, \bibinfo {author} {\bibfnamefont {T.}~\bibnamefont {Hazard}}, \bibinfo
  {author} {\bibfnamefont {C.}~\bibnamefont {Payette}}, \bibinfo {author}
  {\bibfnamefont {K.}~\bibnamefont {Wang}}, \bibinfo {author} {\bibfnamefont
  {D.}~\bibnamefont {Zajac}}, \bibinfo {author} {\bibfnamefont
  {J.}~\bibnamefont {Cady}}, \ and\ \bibinfo {author} {\bibfnamefont {J.~R.}\
  \bibnamefont {Petta}},\ }\href@noop {} {\bibfield  {journal} {\bibinfo
  {journal} {Physical Review B}\ }\textbf {\bibinfo {volume} {92}},\ \bibinfo
  {pages} {035304} (\bibinfo {year} {2015})}\BibitemShut {NoStop}%
\bibitem [{\citenamefont {Laroche}\ \emph {et~al.}(2015)\citenamefont
  {Laroche}, \citenamefont {Huang}, \citenamefont {Nielsen}, \citenamefont
  {Chuang}, \citenamefont {Li}, \citenamefont {Liu},\ and\ \citenamefont
  {Lu}}]{laroche_scattering_2015}%
  \BibitemOpen
  \bibfield  {author} {\bibinfo {author} {\bibfnamefont {D.}~\bibnamefont
  {Laroche}}, \bibinfo {author} {\bibfnamefont {S.-H.}\ \bibnamefont {Huang}},
  \bibinfo {author} {\bibfnamefont {E.}~\bibnamefont {Nielsen}}, \bibinfo
  {author} {\bibfnamefont {Y.}~\bibnamefont {Chuang}}, \bibinfo {author}
  {\bibfnamefont {J.-Y.}\ \bibnamefont {Li}}, \bibinfo {author} {\bibfnamefont
  {C.~W.}\ \bibnamefont {Liu}}, \ and\ \bibinfo {author} {\bibfnamefont
  {T.~M.}\ \bibnamefont {Lu}},\ }\href {\doibase 10.1063/1.4933026} {\bibfield
  {journal} {\bibinfo  {journal} {AIP Advances}\ }\textbf {\bibinfo {volume}
  {5}},\ \bibinfo {pages} {107106} (\bibinfo {year} {2015})}\BibitemShut
  {NoStop}%
\bibitem [{\citenamefont {Friesen}\ \emph {et~al.}(2007)\citenamefont
  {Friesen}, \citenamefont {Chutia}, \citenamefont {Tahan},\ and\ \citenamefont
  {Coppersmith}}]{Friesen2007ValleyWells}%
  \BibitemOpen
  \bibfield  {author} {\bibinfo {author} {\bibfnamefont {M.}~\bibnamefont
  {Friesen}}, \bibinfo {author} {\bibfnamefont {S.}~\bibnamefont {Chutia}},
  \bibinfo {author} {\bibfnamefont {C.}~\bibnamefont {Tahan}}, \ and\ \bibinfo
  {author} {\bibfnamefont {S.~N.}\ \bibnamefont {Coppersmith}},\ }\href
  {\doibase 10.1103/PhysRevB.75.115318} {\bibfield  {journal} {\bibinfo
  {journal} {Physical Review B}\ }\textbf {\bibinfo {volume} {75}},\ \bibinfo
  {pages} {115318} (\bibinfo {year} {2007})}\BibitemShut {NoStop}%
\bibitem [{\citenamefont {Paquelet~Wuetz}\ \emph
  {et~al.}(2020{\natexlab{b}})\citenamefont {Paquelet~Wuetz}, \citenamefont
  {Losert}, \citenamefont {Tosato}, \citenamefont {Lodari}, \citenamefont
  {Bavdaz}, \citenamefont {Stehouwer}, \citenamefont {Amin}, \citenamefont
  {Clarke}, \citenamefont {Coppersmith}, \citenamefont {Sammak}, \citenamefont
  {Veldhorst}, \citenamefont {Friesen},\ and\ \citenamefont
  {Scappucci}}]{wuetz2020effect}%
  \BibitemOpen
  \bibfield  {author} {\bibinfo {author} {\bibfnamefont {B.}~\bibnamefont
  {Paquelet~Wuetz}}, \bibinfo {author} {\bibfnamefont {M.~P.}\ \bibnamefont
  {Losert}}, \bibinfo {author} {\bibfnamefont {A.}~\bibnamefont {Tosato}},
  \bibinfo {author} {\bibfnamefont {M.}~\bibnamefont {Lodari}}, \bibinfo
  {author} {\bibfnamefont {P.~L.}\ \bibnamefont {Bavdaz}}, \bibinfo {author}
  {\bibfnamefont {L.}~\bibnamefont {Stehouwer}}, \bibinfo {author}
  {\bibfnamefont {P.}~\bibnamefont {Amin}}, \bibinfo {author} {\bibfnamefont
  {J.~S.}\ \bibnamefont {Clarke}}, \bibinfo {author} {\bibfnamefont {S.~N.}\
  \bibnamefont {Coppersmith}}, \bibinfo {author} {\bibfnamefont
  {A.}~\bibnamefont {Sammak}}, \bibinfo {author} {\bibfnamefont
  {M.}~\bibnamefont {Veldhorst}}, \bibinfo {author} {\bibfnamefont
  {M.}~\bibnamefont {Friesen}}, \ and\ \bibinfo {author} {\bibfnamefont
  {G.}~\bibnamefont {Scappucci}},\ }\href {\doibase
  10.1103/PhysRevLett.125.186801} {\bibfield  {journal} {\bibinfo  {journal}
  {Physical Review Letters}\ }\textbf {\bibinfo {volume} {125}},\ \bibinfo
  {pages} {186801} (\bibinfo {year} {2020}{\natexlab{b}})}\BibitemShut
  {NoStop}%
\bibitem [{\citenamefont {Tracy}\ \emph {et~al.}(2009)\citenamefont {Tracy},
  \citenamefont {Hwang}, \citenamefont {Eng}, \citenamefont {Ten~Eyck},
  \citenamefont {Nordberg}, \citenamefont {Childs}, \citenamefont {Carroll},
  \citenamefont {Lilly},\ and\ \citenamefont
  {Das~Sarma}}]{Tracy2009ObservationMOSFET}%
  \BibitemOpen
  \bibfield  {author} {\bibinfo {author} {\bibfnamefont {L.~A.}\ \bibnamefont
  {Tracy}}, \bibinfo {author} {\bibfnamefont {E.~H.}\ \bibnamefont {Hwang}},
  \bibinfo {author} {\bibfnamefont {K.}~\bibnamefont {Eng}}, \bibinfo {author}
  {\bibfnamefont {G.~A.}\ \bibnamefont {Ten~Eyck}}, \bibinfo {author}
  {\bibfnamefont {E.~P.}\ \bibnamefont {Nordberg}}, \bibinfo {author}
  {\bibfnamefont {K.}~\bibnamefont {Childs}}, \bibinfo {author} {\bibfnamefont
  {M.~S.}\ \bibnamefont {Carroll}}, \bibinfo {author} {\bibfnamefont {M.~P.}\
  \bibnamefont {Lilly}}, \ and\ \bibinfo {author} {\bibfnamefont
  {S.}~\bibnamefont {Das~Sarma}},\ }\href {\doibase 10.1103/PhysRevB.79.235307}
  {\bibfield  {journal} {\bibinfo  {journal} {Physical Review B}\ }\textbf
  {\bibinfo {volume} {79}},\ \bibinfo {pages} {235307} (\bibinfo {year}
  {2009})}\BibitemShut {NoStop}%
\bibitem [{\citenamefont {Das~Sarma}\ and\ \citenamefont
  {Stern}(1985)}]{das_sarma_single-particle_1985}%
  \BibitemOpen
  \bibfield  {author} {\bibinfo {author} {\bibfnamefont {S.}~\bibnamefont
  {Das~Sarma}}\ and\ \bibinfo {author} {\bibfnamefont {F.}~\bibnamefont
  {Stern}},\ }\href {\doibase 10.1103/PhysRevB.32.8442} {\bibfield  {journal}
  {\bibinfo  {journal} {Physical Review B}\ }\textbf {\bibinfo {volume} {32}},\
  \bibinfo {pages} {8442} (\bibinfo {year} {1985})}\BibitemShut {NoStop}%
\bibitem [{\citenamefont {Qian}\ \emph {et~al.}(2017)\citenamefont {Qian},
  \citenamefont {Nakamura}, \citenamefont {Fallahi}, \citenamefont {Gardner},
  \citenamefont {Watson}, \citenamefont {L{\"u}scher}, \citenamefont {Folk},
  \citenamefont {Csathy},\ and\ \citenamefont {Manfra}}]{qian2017quantum}%
  \BibitemOpen
  \bibfield  {author} {\bibinfo {author} {\bibfnamefont {Q.}~\bibnamefont
  {Qian}}, \bibinfo {author} {\bibfnamefont {J.}~\bibnamefont {Nakamura}},
  \bibinfo {author} {\bibfnamefont {S.}~\bibnamefont {Fallahi}}, \bibinfo
  {author} {\bibfnamefont {G.~C.}\ \bibnamefont {Gardner}}, \bibinfo {author}
  {\bibfnamefont {J.~D.}\ \bibnamefont {Watson}}, \bibinfo {author}
  {\bibfnamefont {S.}~\bibnamefont {L{\"u}scher}}, \bibinfo {author}
  {\bibfnamefont {J.~A.}\ \bibnamefont {Folk}}, \bibinfo {author}
  {\bibfnamefont {G.~A.}\ \bibnamefont {Csathy}}, \ and\ \bibinfo {author}
  {\bibfnamefont {M.~J.}\ \bibnamefont {Manfra}},\ }\href@noop {} {\bibfield
  {journal} {\bibinfo  {journal} {Physical Review B}\ }\textbf {\bibinfo
  {volume} {96}},\ \bibinfo {pages} {035309} (\bibinfo {year}
  {2017})}\BibitemShut {NoStop}%
\bibitem [{\citenamefont {Coleridge}(1991)}]{coleridge1991small}%
  \BibitemOpen
  \bibfield  {author} {\bibinfo {author} {\bibfnamefont {P.}~\bibnamefont
  {Coleridge}},\ }\href@noop {} {\bibfield  {journal} {\bibinfo  {journal}
  {Physical Review B}\ }\textbf {\bibinfo {volume} {44}},\ \bibinfo {pages}
  {3793} (\bibinfo {year} {1991})}\BibitemShut {NoStop}%
\end{thebibliography}%

\end{document}